\documentclass[twocolumn]{aastex7}
\usepackage[version=4]{mhchem}

\begin{document}

 \title{Limits on forming coreless terrestrial worlds in the TRAPPIST-1 system}

\author[orcid=0000-0002-9498-851X,sname='Huang']{Dongyang Huang}
\affiliation{SKLab-DeepMinE, MOEKLab-OBCE, School of Earth and Space Sciences, Peking University, 100871 Beijing, China}
\email[show]{dhuang@pku.edu.cn}
\affiliation{Research Institute of Extraterrestrial Material at PKU (RIEMPKU)}

\author[orcid=0000-0001-6110-4610,gname=Caroline, sname='Dorn']{Caroline Dorn} 
\affiliation{Institute for Particle Physics and Astrophysics, ETH Zürich, 8093 Zürich, Switzerland}
\email[show]{dornc@phys.ethz.ch}

\begin{abstract}

With seven temperate Earth-sized planets revolving around an ultracool red dwarf, the nearby TRAPPIST-1 system offers a unique opportunity to verify models of exoplanet composition, differentiation, and interior structure. 
In particular, the low bulk densities of the TRAPPIST-1 planets, compared to terrestrial planets in our solar system, require either substantial amount of volatiles to be present or a corefree scenario where the metallic core is fully oxidised.
Here, using an updated metal–silicate partitioning model, we show that during core–mantle differentiation oxygen becomes more siderophile (iron-loving) implying larger planet radii. 
For the seven TRAPPIST-1 planets, however, we find that they are not sufficiently large to oxidise all the iron in the core, if they differentiate from an Earth-like composition.
Oxygen partitioning in rocky worlds precludes coreless planets up to $\sim$4 M$_\oplus$.
The observed density deficit in the TRAPPIST-1 planets, and more generally in M dwarf systems if confirmed by future observations, may be explained by system-dependent element budgets during planet formation, which are intrinsically linked to their stellar metallicity.

\end{abstract}



\section{Introduction}

TRAPPIST-1 is a low-mass ($\sim$0.09 M$_\odot$) M8 dwarf star, at a distance of $\sim$12 parsecs from Earth, with seven Earth-sized transiting rocky planets located within or near the edge of its habitable zone \citep{Gillon2017}.
As an important target for both detecting the atmospheric signatures of potentially habitable worlds and for testing theories of formation, migration, and interior structure for terrestrial planets around M dwarfs, the system has received much attention since its discovery \citep{Luger2017,Quarles2017,Grimm2018,Agol2021,Krissansen-Totton2022,Piaulet-Ghorayeb2025}.
In particular, based on transit photometry and transit-timing variations (TTVs), the planetary radii and masses are constrained to $\sim$0.76–1.1 R$_\oplus$ and $\sim$0.3–1.4 M$_\oplus$, both to a precision of less than 5\% \citep{Grimm2018,Agol2021}, establishing the Earth-like nature in all seven planets.
All seven TRAPPIST-1 planets follow an iso-chemical trend in the mass–radius diagram \citep{Agol2021} with densities slightly lower than an Earth-like composition; we refer to this as the `density deficit'.
Like compositional models designed to account for the mass–radius relationships for exoplanets \citep{Seager2007,Dorn2015,Zeng2019,Noack2020,Luque2022,Spaargaren2023}, the interpretation of the nature of this density deficit suffers from the degeneracy between 
rocky components (including an iron core, a silicate mantle) and volatile-rich components \citep{Unterborn2018,Agol2021}.
For instance, the estimated amount of water, the most popular volatile species, from varying models ranges from dry, or Earth-like (0.001–5 wt\%) \citep{Dorn2018,Agol2021} to water-rich worlds (15–$\geq$50 wt\%) \citep{Quarles2017,Dorn2018,Unterborn2018} for the TRAPPIST-1 planets.
Alternative solutions to the density deficit include varying core–mantle fraction (CMF) or Fe/Mg ratio \citep{Valencia2007,Agol2021}, or an extreme core-free scenario where all the iron in the core is oxidised and mixed with the mantle \citep{ElkinsTanton2008,Agol2021}.

From a planet formation perspective, it is possible that the TRAPPIST-1 planets formed beyond the snow line and migrated inward \cite[e.g.][]{Schoonenberg2019}, implying that significant amounts of water are possible. However, \cite{Chen2025} argued that rocky planets in compact M-dwarf systems form depleted in volatiles,  allowing for only few Earth oceans accreted from the TRAPPIST-1 planets. The sustainability of atmospheres or the presence of surface oceans is thus limited. This is also supported by the fact that any atmosphere may be subject to efficient loss leaving behind negligibly thin atmospheres \citep{van2025habitable, pass2025receding} with no effect on mass-radius relationships.  

Recent JWST observations of the TRAPPIST-1 system indicate that the two innermost planets, TRAPPIST-1 b and c, lack substantial atmospheres \citep{ih_constraining_2023,greene_thermal_2023,Lim2023,Lincowski2023,Zieba2023,Ducrot2025,Radica2025}: their thermal phase curves show poor heat redistribution, and transmission spectra are featureless, consistent with bare-rock surfaces. This aligns with theoretical expectations that intense stellar irradiation ($\sim$10$^3$–10$^4$ times XUV fluxes that of the modern Earth) would have driven significant atmospheric loss \citep{Fleming_2020,JiXuan_2025,Chatterjee2025}. TRAPPIST-1 d, receiving more moderate flux, remains a marginal case—models suggest roughly a 70\% probability of retaining some atmosphere \citep{krissansen2023implications,gialluca2024implications}, though others are less optimistic \citep{van2025habitable}. JWST observations rule out clear hydrogen-rich or dense \ce{CO2}- and \ce{CH4}-rich atmospheres ($>$ 3$\sigma$ confidence), instead favoring either an extremely tenuous or aerosol-laden atmosphere, or none at all \citep{Piaulet-Ghorayeb2025}. Previous Hubble and Spitzer data were only sensitive enough to exclude clear, low–mean-molecular-weight atmospheres, leaving high–mean-molecular-weight or hazy compositions unconstrained. Overall, spectroscopic observations are consistent with the absence of any thick atmospheres on the innermost three planets.

The absence of a thick atmosphere reduces the degeneracy in possible interior models to the composition and layer thicknesses of a core and a mantle.
Interior structure models often rely on first-order assumptions. Commonly, planets are modeled as a set of separate layers of distinct compositions or phases. This limitation, among other factors, adds to the degeneracy of interior models and limits their accuracy.
The reason is that commonly used interior models neglect core–mantle partitioning of volatiles (such as H, C and N) which is prevalent during planet formation \citep{li_earths_2020,Fischer2020,Huang2024}. A recent work calculated global water distribution between exoplanet cores and mantles using a thermodynamic integration method \citep{Luo2024}. The majority of the accreted water of a planet is found to be stored deep in the metal core, reducing the available surface water by orders of magnitudes. This finding was confirmed by chemical equilibration models to be valid also for sub-Neptunes \citep{Werlen2025}. Water is efficiently lost to the deep interior.
Volatile partitioning in deep interiors is essential for accurate models, as surface reservoirs alone can differ by orders of magnitude from the total bulk inventory \cite[e.g.][]{Dorn2021,Luo2024}. Bulk volatile abundances are what link a planet to its primordial composition, providing the key to reconstructing its formation pathway and original location within the protoplanetary disk, as is the case with the provenance of the terrestrial C, N and S budgets \citep{Huang2024}.

For modeling the interiors of the TRAPPIST-1 planets, we are agnostic about the volatile-rich terrestrial planets, as the partitioning behaviour of most volatile species remains undetermined at pressures (100–1000 GPa) and concentration level ($>$ 10 wt\%) relevant to exoplanet interiors. Furthermore, we test different assumptions for planetary core–mantle fraction (CMF). We use the canonical assumption that the TRAPPIST-1 planets follow an Earth-like CMF; and also test a CMF that is based on TRAPPIST-1 stellar metallicity estimates \citep{gillon_temperate_2016} and its interpretation regarding bulk Fe/Mg \citep{unterborn_inward_2018}.
Here we propose to explore the third scenario, i.e. the coreless case, by devising a thermodynamic model for the partitioning of oxygen between cores and mantles of Earth-sized exoplanets. 
After testing the viability of the coreless scenario in the TRAPPIST-1 system, which provides a rare statistics with seven terrestrial planets within a single system, we extend our discussions to potential implications for atmospheric signatures as a consequence of core–mantle–atmosphere equilibration.

\section{Methods}

\subsection{Thermodynamic model: oxygen partitioning between the core and mantle of rocky planets}
Oxygen (O) is the third most abundant element in the solar system after H and He \citep{Lodders2003}, and the most abundant element, apart from Fe, in Earth \citep{Wang2018}. 
Consequently, the partitioning of O between Fe metal and silicate determines the extent to which the metallic core is oxidised.

Oxygen partitioning to iron was previously considered limited based on experiments performed at relatively low pressures up to Earth's upper mantle conditions. Thanks to recent advances in laser-heated diamond anvil cell experiments \citep[e.g.][]{Bouhifd2011,Siebert2012}, numerous new data became available in the past decade, where O was found to dissolve into Fe in substantial amounts at Earth's core–mantle boundary pressure \citep{Siebert2013,Badro2016,Huang2018,Huang2020,Huang2021,Huang2024}.
This makes it possible to device a new thermodynamic model for oxygen. 
The partitioning of O is accompanied by the dissolution of oxides from the silicate into the metal, which, in terms of major rock-forming cations (Si, Mg, and Fe), can be described by
\begin{equation}
	SiO_2^{silicate} \rightleftharpoons Si^{metal} + 2O^{metal},
\end{equation}
\begin{equation}
	MgO^{silicate} \rightleftharpoons Mg^{metal} + O^{metal},
\end{equation}
and
\begin{equation}
	FeO^{silicate} \rightleftharpoons Fe^{metal} + O^{metal},
	\label{eq3}
\end{equation}
whose universal form is therefore
\begin{equation}
	[O^{2-}]^{silicate} \rightleftharpoons [O^{0}]^{metal}.
	\label{eq4}
\end{equation}

The equilibrium constant $K$ of reaction \ref{eq4}, defined in terms of activities ($a_i$) of the components, their mole fractions $c_i$ and activity coefficients $\gamma_i$ in relevant phases, is given by 
\begin{equation}
	K=\frac{a_{O^0}^{metal}}{a_{O^{2-}}^{silicate}}=\frac{c_{O^0}^{metal}\cdot \gamma_{O^0}^{metal}}{c_{O^{2-}}^{silicate}\cdot \gamma_{O^{2-}}^{silicate}}.
	\label{eq_k}
\end{equation}
Assuming only temperature dependence for $\gamma_{O^0}^{metal}$, the term can be approximated as the activity of O in liquid Fe in infinite dilution, of which we take the form provided by \cite{Badro2015}. $\gamma_{O^{2-}}^{silicate}$ is unknown, and is assumed to be unity, as O is by number the most abundant species in the silicate.
For experimentally determined concentrations of oxygen ($c_i$ in metal and silicate phases), we use the compilation by \cite{Huang2024}. Their ratio is by definition the partition coefficient $D$ of oxygen between metal and silicate
\begin{equation}
	D_O = c_{O^0}^{metal} / c_{O^{2-}}^{silicate}.
	\label{eq6}
\end{equation}

Based on thermodynamic relations, equilibrium constant $K$ can be expressed as a function of P (in GPa) and T (in K) \citep{Huang2021,Huang2024}; by fitting the experimental data and by rearranging Eqs. \ref{eq_k}–\ref{eq6}, one obtains
\begin{equation}
	D_O = \frac{10^{2.066(\pm0.280) - 12378(\pm697)/T + 37.76(\pm8.97)\cdot P/T}}{exp(4.29 - 16500/T)}.
	\label{eq7}
\end{equation}

\subsection{Interior models coupled with oxygen partitioning}

In order to quantify the implications of self-consistent oxygen partitioning for planet interiors, we build upon the models from \citep{Dorn2015, dorn_generalized_2017} with latest updates in \citep{Luo2024}. For this application, the planets of interest are rocky worlds only. 

We consider a core made of Fe and O, or equivalently Fe and FeO in terms of composition. For solid Fe, we use the equations of state for hexagonal close packed (hcp) iron \citep{hakim_new_2018,miozzi_new_2020}. For liquid iron, we use \citet{Luo2024}. The core thermal profile is assumed to be adiabatic throughout the core. At the core-mantle boundary (CMB), there is a temperature jump as the core can be hotter than the mantle due to the residual heat released during core formation. We follow \citet{stixrude_melting_2014} and add a temperature jump at the CMB temperature such that the temperature at the top of the core is at least as high as the melting temperature of the silicates. 

The amount of FeO in the core is calculated from the partitioning coefficient of oxygen following equation \ref{eq6}. As the mass fraction of FeO can vary between 0 and 1, corresponding oxygen mass fraction ranges between zero and 0.22. For the equilibration, we use a pressure that equals 30$\%$ of the pressure at the core-mantle-boundary (CMB) and a temperature equal to the dry melting temperature of silicates at this corresponding pressure plus 1000 K.

The mantle is assumed to be composed of three major constituents, i.e., MgO, SiO$_2$, FeO. For the solid mantle, we use the thermodynamical model \verb|Perple_X| \citet{connolly2009geodynamic}, to compute stable mineralogy and density for a given composition, pressure, and temperature, employing the database of \citet{stixrude_thermal_2022}. For pressures higher than $\sim125$~GPa, we define stable minerals \textit{a priori} and use their respective equation of states from various sources \citep{fischer_equation_2011, faik_equation_2018, luo_equation_2023, hemley_constraints_1992, musella_physical_2019}. For the liquid mantle, we calculate its density assuming an ideal mixture of main components (Mg$_2$SiO$_4$,SiO$_2$,FeO) \citep{stewart_shock_2020,faik_equation_2018,melosh2007hydrocode,ichikawa_ab_2020} and add them using the additive volume law. Note that we use Mg$_2$SiO$_4$ melt instead of MgO melt since the data for forsterite has been recently updated for the high-pressure temperature regime, which is not available for MgO to our knowledge. The mantle is assumed to be fully adiabatic. 

\section{Results}
\subsection{Pressure and temperature dependence of O partitioning: Insights from recent LHDAC experiments}

\begin{figure*}
	\centering
	\includegraphics[width=1.\textwidth, trim=2.3cm 0cm 2.9cm 0cm, clip]{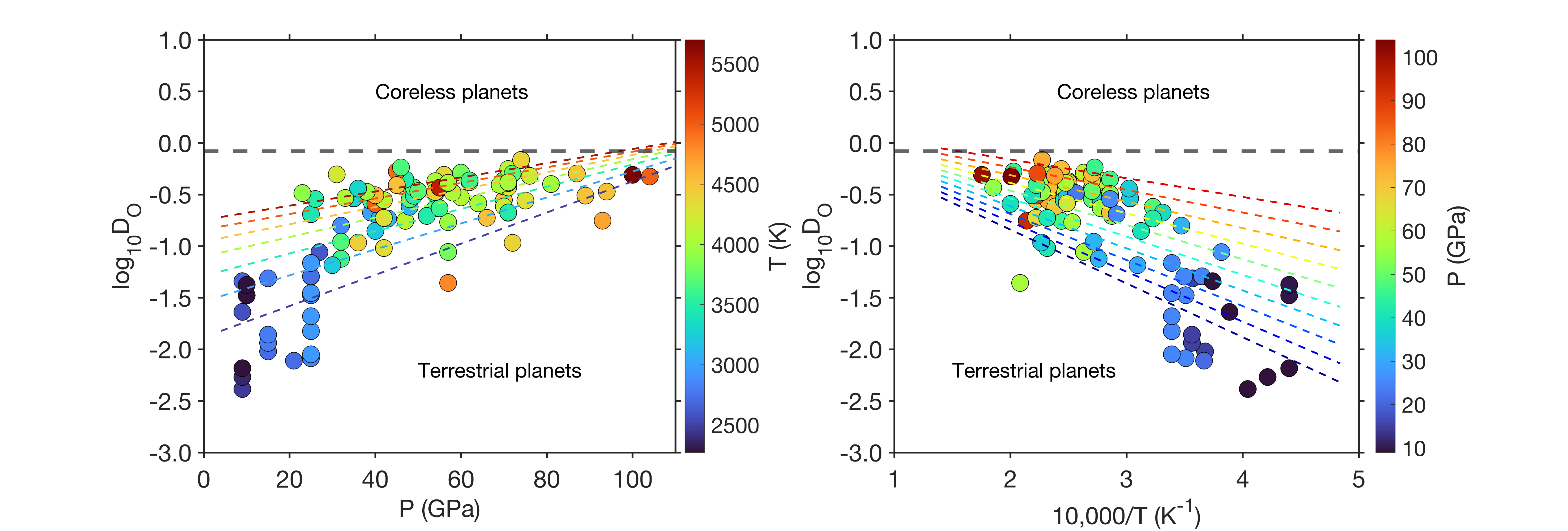}    
	\caption{Core–mantle partition coefficient of oxygen as a function of pressure and temperature. Left: Experimental data for metal–silicate partitioning of O obtained at high pressures up to 104 GPa. Horizontal dashed line located at $D_O$=0.83 indicates full oxidation of the core (see the main text), and therefore distinguishes `terrestrial' from `coreless' planets. It is clear that, within the investigated pressure range, $D_O$ asymptotically approaches the dashed line, but never reaches the core-free regime, albeit higher pressures and temperatures promote the sequestration of O into the metallic core, i.e. larger $D_O$. Right: The same dataset for O partitioning plotted as a function of reciprocal temperature. Dashed lines are fitted results (Eq. \ref{eq7}) showing the dependence of O partitioning on temperature at a given pressure or vice versa.}
	\label{fig:O_model}
\end{figure*}

The most remarkable feature of the O partitioning is its positive correlation to increasing pressure and temperature, which corresponds to more massive planets with deeper magma oceans during their formations. Despite of the scattering nature of the experimental data at a specific pressure, which is somewhat unavoidable as experimental results are affected by other variables, such as temperature and composition, apart from the control variable pressure \citep{Bouhifd2011,Siebert2013,Badro2016,Huang2018,Huang2020,Huang2021,Huang2024}, the overall positive correlation between $D_O$ and pressure (and temperature) is evident (Fig. \ref{fig:O_model}, left panel). At a given pressure, temperature increases $D_O$, partitioning more O into the core; at a given temperature, higher pressure similarly makes O more siderophile (Fig. \ref{fig:O_model}, right panel). 
It is important to distinguish between two sorts of equilibrations occuring at different stages of rocky planets in a molten state: (i) metal–silicate equilibration (i.e. Eq. \ref{eq3}) that takes place during core–mantle differentiation, and (ii) a disproportionation of ferrous iron ($3Fe^{2+} = Fe^{0}+2Fe^{3+}$) in silicate melts \citep{Armstrong2019} subsequent to core–mantle segregation, where the majority of metallic iron had been removed/sequestered into the core. 
Our model focuses on the former, earlier event, treating the equilibration between the bulk core and mantle, rather than the latter, which exclusively deals with silicate melt and pressure-induced disproportionation within it.

Because the molar fraction of O in an Earth-like mantle is invariant with $c_{O^{2-}}^{silicate}=\sim0.6$, given a terrestrial Mg/Si molar ratio of 0.9–1.2, if the core is fully oxidised with the Fe/O molar ratio equals unity, in principle the core-free planets emerge when $D_O=0.83$ (dashed lines in Fig. \ref{fig:O_model}), as a result of sufficiently high pressure core–mantle differentiation. An equally significant observation from these experiments is that, although the partition coefficient of O points to the direction of coreless planets, as $D_O$ asymptotically approaches 0.83, it remains within the `terrestrial' regime with differentiated core and mantle for the pressure regime of interest.

\subsection{Oxygen content of the core increases with planet mass}

\begin{figure*}
	\centering
	\includegraphics[width=1.\textwidth, trim=2.3cm 0cm 2.9cm 0cm, clip]{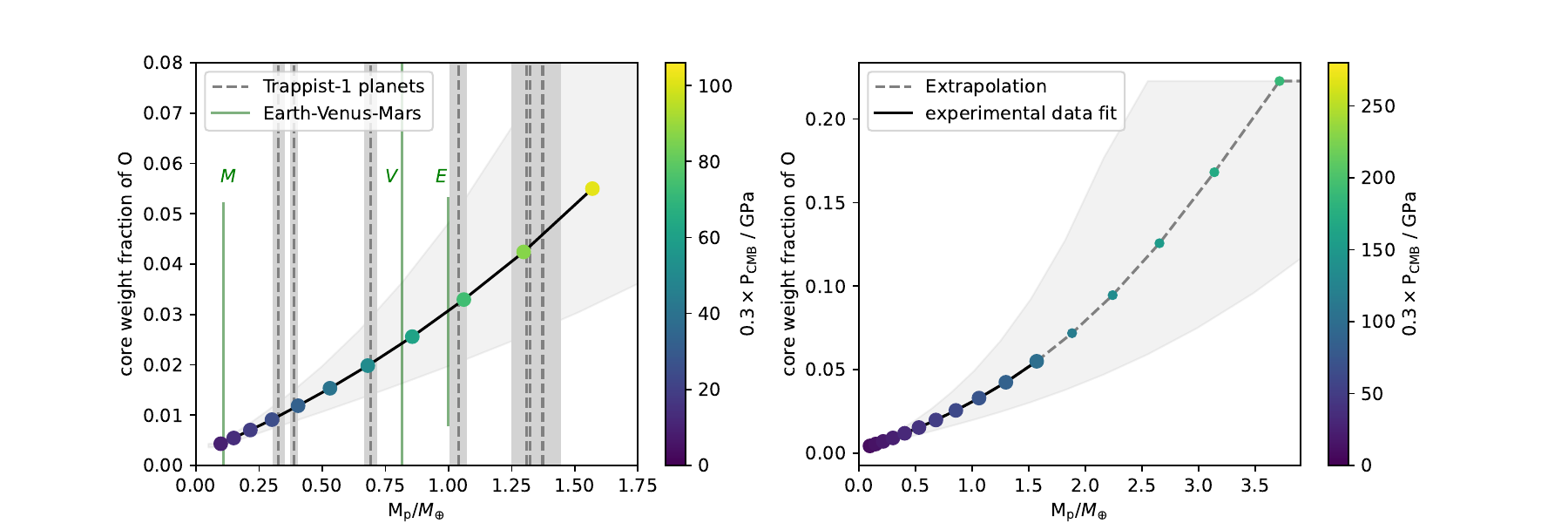}    
	\caption{Equilibrated core oxygen mass fraction as a function of planetary mass.
		Left: Results within the pressure range constrained by experimental data. Calculated core mass fractions are compared with independent estimates for Earth and Mars, and the mass range of the TRAPPIST-1 planets is also indicated.
		Right: Extrapolation of the metal–silicate equilibration model to higher masses and pressures. The oxygen mass fraction approaches an upper limit of 0.22, corresponding to complete oxidation of the core as FeO. In reality, this limit would be approached asymptotically rather than a sharp transition shown at 3.5 M$_\oplus$. Light grey areas illustrate the uncertainty on oxygen partitioning (Eq. \ref{eq7}).}
	\label{fig:P_FeO}
\end{figure*}

We aim to quantify implications of the new  oxygen partitioning model for interior structures of super-Earths.
As the partitioning coefficient of oxygen increases with higher pressure and asymptotically reaches the value of 0.83, we expect a clear dependency of core oxygen content with planet mass. Figure \ref{fig:P_FeO} clearly illustrates that more massive planets have more oxidized cores than smaller planets. As a result, pure iron cores become implausible for large planet masses. Instead, planets beyond $\sim$ 3.5 M$_\oplus$, are expected to have fully oxidized cores as predicted by our oxygen partitioning model. Such planets experience equilibration pressures beyond 200 GPa, where no experimental data is currently available. The experimentally constrained pressure range only reaches $\sim$ 110 GPa (Figure \ref{fig:P_FeO}. Such pressures are relevant for silicate-metal equilibration up to planet masses of $\sim$1.6 M$_\oplus$. This is true under the assumption that the equilibration pressures are 30 \% of the CMB pressure. If the overall equilibration happens deeper in planet mantles, the range of planet masses that is covered by the experimental data would be more restricted to smaller masses. In general, the trend of increasing core oxygen content with planet mass is a robust finding, the actual gradient differs upon assumptions of the equilibration pressure and temperature.

Further, our oxygen partitioning model agrees with independent estimates for Mars, Earth, and Venus, although constraints for the terrestrial planets are limited (Fig. \ref{fig:P_FeO}). The values for the terrestrial planets come from \cite{Huang2023} and \cite{khan_evidence_2023} for Mars, \cite{Tronnes2019} for Venus, and \cite{hirose_light_2021} for Earth.

\subsection{M-R diagram revisited: the TRAPPIST-1 planets}

\begin{figure*}
	\centering
	\includegraphics[width=1.\textwidth, trim=2.3cm 0cm 2.9cm 0cm, clip]{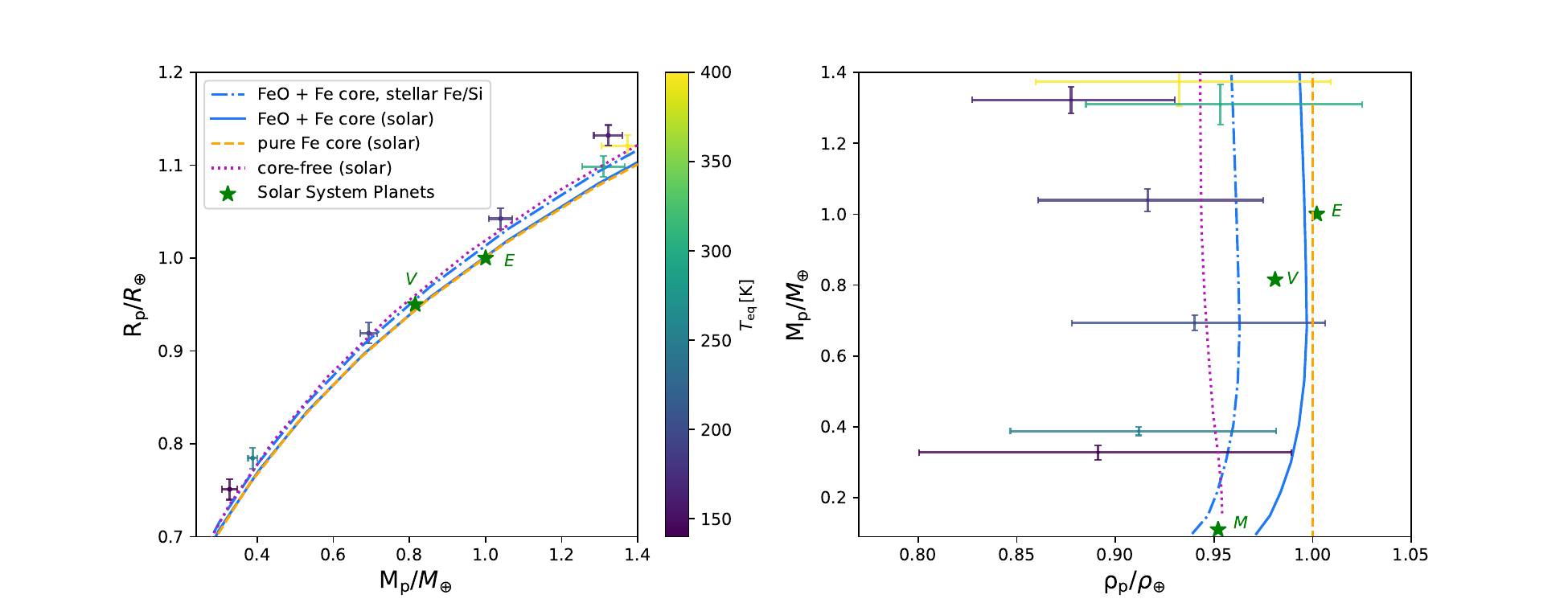}    
	\caption{Left: Mass-radius diagram compared to the TRAPPIST-1 planets. Right: mass-normalized density diagram. Densities are normalized to the nominal model of a pure Fe core with solar Fe/Si abundances (orange line). The metal–silicate-equilibrated curve where oxygen partitioning is self-consistently modeled for solar Fe/Si (blue solid line) plots between the end-members of a pure Fe core (orange line) and core-free model (pink line). When the stellar Fe/Si estimate from \citep{Unterborn2018} is used, the line shifts to lower densities (blue dash-dotted line), reasonably fitting most TRAPPIST planets.}
	\label{fig:MR}
\end{figure*}

It is clear that the TRAPPIST-1 planets span a range of masses that would imply different oxygen core contents ranging from 0.01 to 0.05 wt\%. Can this trend be observed or tested? Figure \ref{fig:MR} shows mass-radius curves that incorporate self-consistently the change in core FeO content with planet mass (blue lines). First, we compare the new self-consistent interior model assuming identical solar refractory (Fe, Si, Mg) abundance: Compared to a standard rocky interior model with a pure iron and an iron-free mantle (orange dashed line), there is little difference over the small mass range of the TRAPPIST-1 planets (orange dashed line is similar to the blue solid line). Only for larger masses $>$ 3.5 M$_\oplus$, when the metal phase are saturated with oxygen, the mass-radius curves (blue solid) will approach the densities for the core-free case (pink dotted line). 

The core-free interior density for solar abundance (pink dotted) is only $5\%$ lower from the pure Fe-core (orange dashed). Interestingly, this small density decrease is just right to fit all TRAPPIST-1 planets which was an intriguing result from \citet{Agol2021}. Of course, there is inherent degeneracy and the TRAPPIST-1 planets can also be explained by lower Fe/Mg bulk ratios or by individual variations in water content from planet to planet. However, the latter scenario is less probable, as spectroscopic observations with JWST have ruled out thick atmospheres for the innermost planets, and reproducing the required planet-by-planet water fractions would demand finely tuned formation pathways. Here, we now add strong constraints from mineral physics to exclude the core-free scenario.

If we allow refractory element ratios to depart from solar, the mass-radius trend will differ. We adopt bulk abundance constraints as suggested by \citet{Unterborn2018} who analyzed F-G-K stars of similar metallicity to TRAPPIST-1 and estimate a median of Fe/Si = $1.49\pm 0.4$ in mass ratios, while fixing the Mg/Si mass ratio to 0.87. Using a similar approach but excluding thick-disc stars, \cite{acuna2021characterisation} estimated a very similar median of Fe/Si = 1.51. For comparison, the solar mass ratio estimates are Fe/Si = 1.69. 
With the adopted abundance from \citet{Unterborn2018}, all planets except for TRAPPIST-1~g ($T_{eq} = 199$K) fall onto the rocky mass-radius curve for equilibrated metal–silicate (blue dash-dotted line) within their 1-$\sigma$ uncertainties. Note that it is the difference in stellar abundances that allows most planets to be fitted within their 1-$\sigma$ error bars, while changes due to metal–silicate equilibration are minor across the range of TRAPPIST-1 planets.

\section{Discussion}

Whether or not FeO cores would actually form from silicate melts early in planet histories remains an open question for planet masses $>$ 3.5 M$_\oplus$. The current experimental constraints and thermodynamic model show the asymptotical trend of O partitioning towards a fully oxidised core (Fig. \ref{fig:O_model}). It appears to be unlikely that there exists a coreless scenario at higher P-T conditions within more massive Earth-like planets, which is different from the water-rich scenario proposed by \cite{ElkinsTanton2008} . To verify whether FeO can differentiate to form a core or not, new high-pressure experiments under expanded P–T–composition space would be necessary in the future (see Luo et al. in prep.).
If verified, experimentally or theoretically, the predicted FeO core, existing when M\textsubscript{P} $>$ 3.5 M$_\oplus$(Fig. \ref{fig:P_FeO}), could readily become miscible with silicate melts, potentially refuting the conventional assumption that interiors of large rocky planets have to differentiate into a layered structure. 

Similarly, based on mixing properties in the binary MgSiO$_3$–Fe (in fact MgO–Fe), MgSiO$_3$–H$_2$, and Fe-H$_2$ systems, \cite{young2025} concluded that MgSiO$_3$, Fe, and H$_2$ become fully miscible in deep interiors of molten cores (anything beneath the outer envelope, i.e. including silicate mantles and metallic cores) of sub-Neptunes or their super-Earth descendants. Considering the enhanced oxygen partitioning at high P-T conditions, FeO instead of pure Fe, is a more likely end-member and may enable full miscibility at even shallower depths. 

The partitioning of oxygen into the metal phase has important implications for atmospheres of sub-Neptunes and super-Earths.
Firstly, the oxidation of metallic iron limits the amount of O available to oxidize H. 
As a result, the amount of atmospheric \ce{H2O} that can be outgassed would decrease with increasing planet mass.
This is broadly consistent with \cite{schlichting_chemical_2022}, where water concentrations in silicate mantles and envelopes of super-Earths are significantly lower if one takes into account the chemical equilibration between the core, mantle, and atmosphere.
Secondly, as noted by \cite{Werlen2025}, parameters such as planet mass, relative mass fractions of core–mantle–atmosphere, all strongly influence the atmospheric C/O ratio. In a closed system with fixed initial amounts of  C and O \citep{schaefer_redox_2017},  if the metal–silicate (metallic core–silicate magma ocean) equilibration is considered, the partitioning of both elements into iron would determine the amounts of C and O available in the magma ocean, which would eventually modify the observed atmospheric C/O ratios owing to magma ocean-atmosphere interactions \citep{Seo2024,Lichtenberg2025,Nixon2025,Bower_2025}.

A caveat with the current model of O partitioning is that it is established on the basis of experimental data on relatively `dry' terrestrial materials \citep{Huang2018,Huang2020,Huang2021,Huang2024}, and is hence applicable to rocky interiors or lava planets in the regime of Henry's law; its applicability to ocean planets \citep{Kuchner2003,Leger2004}, formed beyond the snow line and containing 10s wt\% level of water, may be limited and requires further investigations by the mineral physics community in the future.
As a step towards a more realistic scenario, future studies should address interactions among volatiles (H, C, N and O) when metal and silicate react with each other \citep[e.g.][]{Huang2024}, and how they affect the partitioning and thereby atmospheric compositions.

\section{Conclusions}

We present a new metal–silicate equilibration model that predicts the oxygen content of the metal phase under given pressure–temperature conditions. The model is constrained by experimental data spanning pressures from a few GPa up to 104 GPa \citep{Huang2024}. Our results show a clear trend of increasing oxygen incorporation with increasing planetary mass, under the assumption that equilibration pressure scales with mass. Although extrapolation is required beyond ~2 M$_\oplus$, where equilibration pressures exceed experimental limits, the model predicts that pure Fe cores become impossible above a critical planet mass. Specifically, the extrapolation suggests this threshold occurs near 3.5 M$_\oplus$, though the exact value depends on the assumptions adopted here. The key result is that as the planet approaches this critical mass, its metallic core will become saturated with oxygen.

The implications for the TRAPPIST-1 planets are that coreless scenarios can be ruled out, and that the oxygen content among their cores likely ranges from 0.01 to 0.05 wt\%. Previously, the possible interior configurations included coreless models, variations in bulk Fe/Mg (or Fe/Si) ratios, and volatile-rich compositions. However, given recent spectroscopic observations from JWST and evolutionary constraints on water loss, we are left with a single plausible explanation: the Fe/Mg (or Fe/Si) ratios are sub-solar. This interpretation is consistent with previous estimates of the stellar composition \citep{unterborn_inward_2018, dorn_interior_2018, acuna2021characterisation}.
The fact that super-Earth metal phases become oxygen saturated for high mass planets has possible implications for any atmosphere that can degas from an early mantle magma ocean, including the observed atmospheric H$_2$O and C/O ratio.

\begin{acknowledgments}
D.H. acknowledges supports from PKU startup grant (No. 7100604887) and the National Natural Science Foundation of China (NSFC) under grant 8206100810.
C.D. gratefully acknowledges support from the Swiss National Science Foundation under grant TMSGI2\_211313. This work has been carried out within the framework of the NCCR PlanetS supported by the Swiss National Science Foundation under grant 51NF40\_205606. 
\end{acknowledgments}

\begin{contribution}

All authors contributed equally to the manuscript.


\end{contribution}

\bibliography{references01,references02}{}

@article{Huang2018,
author = {Huang, Dongyang and Badro, James},
doi = {10.2138/am-2018-6651},
issn = {19453027},
journal = {American Mineralogist},
month = {oct},
number = {10},
pages = {1707--1710},
publisher = {De Gruyter Open Ltd},
title = {{Fe-Ni ideality during core formation on Earth}},
volume = {103},
year = {2018}
}

@article{Huang2023,
author = {Huang, D and Li, Y and Khan, A and Sossi, P and Giardini, D and Murakami, M},
doi = {10.1029/2022GL102271},
issn = {0094-8276},
journal = {Geophysical Research Letters},
month = {mar},
number = {6},
pages = {1--11},
title = {{Thermoelastic Properties of Liquid Fe‐Rich Alloys Under Martian Core Conditions}},
url = {https://onlinelibrary.wiley.com/doi/10.1029/2022GL102271 https://agupubs.onlinelibrary.wiley.com/doi/10.1029/2022GL102271},
volume = {50},
year = {2023}
}

@article{Chen2025,
author = {Chen, Howard and Clement, Matthew S. and Wang, Le``Chris” and Gu, Jesse T.},
doi = {10.3847/2041-8213/adf282},
issn = {2041-8205},
journal = {The Astrophysical Journal Letters},
number = {1},
pages = {L11},
publisher = {IOP Publishing},
title = {{Born Dry or Born Wet? A Palette of Water Growth Histories in TRAPPIST-1 Analogs and Compact Planetary Systems}},
url = {http://dx.doi.org/10.3847/2041-8213/adf282},
volume = {991},
year = {2025}
}

@article{Schoonenberg2019,
archivePrefix = {arXiv},
arxivId = {1906.00669},
author = {Schoonenberg, Djoeke and Liu, Beibei and Ormel, Chris W. and Dorn, Caroline},
doi = {10.1051/0004-6361/201935607},
eprint = {1906.00669},
issn = {14320746},
journal = {Astronomy and Astrophysics},
pages = {1--15},
title = {{Pebble-driven planet formation for TRAPPIST-1 and other compact systems}},
volume = {627},
year = {2019}
}

@article{Lim2023,
archivePrefix = {arXiv},
arxivId = {2309.07047},
author = {Lim, Olivia and Benneke, Bj{\"{o}}rn and Doyon, Ren{\'{e}} and MacDonald, Ryan J. and Piaulet, Caroline and Artigau, {\'{E}}tienne and Coulombe, Louis-Philippe and Radica, Michael and L'Heureux, Alexandrine and Albert, Lo{\"{i}}c and Rackham, Benjamin V. and de Wit, Julien and Salhi, Salma and Roy, Pierre-Alexis and Flagg, Laura and Fournier-Tondreau, Marylou and Taylor, Jake and Cook, Neil J. and Lafreni{\`{e}}re, David and Cowan, Nicolas B. and Kaltenegger, Lisa and Rowe, Jason F. and Espinoza, N{\'{e}}stor and Dang, Lisa and Darveau-Bernier, Antoine},
doi = {10.3847/2041-8213/acf7c4},
eprint = {2309.07047},
issn = {2041-8205},
journal = {The Astrophysical Journal Letters},
month = {sep},
number = {1},
pages = {L22},
title = {{Atmospheric Reconnaissance of TRAPPIST-1 b with JWST/NIRISS: Evidence for Strong Stellar Contamination in the Transmission Spectra}},
url = {https://iopscience.iop.org/article/10.3847/2041-8213/acf7c4},
volume = {955},
year = {2023}
}

@article{Radica2025,
archivePrefix = {arXiv},
arxivId = {2409.19333},
author = {Radica, Michael and Piaulet-Ghorayeb, Caroline and Taylor, Jake and Coulombe, Louis-Philippe and Benneke, Bj{\"{o}}rn and Albert, Loic and Artigau, {\'{E}}tienne and Cowan, Nicolas B. and Doyon, Ren{\'{e}} and Lafreni{\`{e}}re, David and L'Heureux, Alexandrine and Lim, Olivia},
doi = {10.3847/2041-8213/ada381},
eprint = {2409.19333},
issn = {2041-8205},
journal = {The Astrophysical Journal Letters},
number = {1},
pages = {L5},
publisher = {IOP Publishing},
title = {{Promise and Peril: Stellar Contamination and Strict Limits on the Atmosphere Composition of TRAPPIST-1 c from JWST NIRISS Transmission Spectra}},
volume = {979},
year = {2025}
}

@article{Lincowski2023,
archivePrefix = {arXiv},
arxivId = {2308.05899},
author = {Lincowski, Andrew P. and Meadows, Victoria S. and Zieba, Sebastian and Kreidberg, Laura and Morley, Caroline and Gillon, Micha{\"{e}}l and Selsis, Franck and Agol, Eric and Bolmont, Emeline and Ducrot, Elsa and Hu, Renyu and Koll, Daniel D. B. and Lyu, Xintong and Mandell, Avi and Suissa, Gabrielle and Tamburo, Patrick},
doi = {10.3847/2041-8213/acee02},
eprint = {2308.05899},
issn = {2041-8205},
journal = {The Astrophysical Journal Letters},
number = {1},
pages = {L7},
title = {{Potential Atmospheric Compositions of TRAPPIST-1 c Constrained by JWST/MIRI Observations at 15 $\mu$m}},
volume = {955},
year = {2023}
}

@article{Zieba2023,
archivePrefix = {arXiv},
arxivId = {2306.10150},
author = {Zieba, Sebastian and Kreidberg, Laura and Ducrot, Elsa and Gillon, Micha{\"{e}}l and Morley, Caroline and Schaefer, Laura and Tamburo, Patrick and Koll, Daniel D.B. and Lyu, Xintong and Acu{\~{n}}a, Lorena and Agol, Eric and Iyer, Aishwarya R. and Hu, Renyu and Lincowski, Andrew P. and Meadows, Victoria S. and Selsis, Franck and Bolmont, Emeline and Mandell, Avi M. and Suissa, Gabrielle},
doi = {10.1038/s41586-023-06232-z},
eprint = {2306.10150},
issn = {14764687},
journal = {Nature},
number = {7975},
pages = {746--749},
pmid = {37337068},
publisher = {Springer US},
title = {{No thick carbon dioxide atmosphere on the rocky exoplanet TRAPPIST-1 c}},
volume = {620},
year = {2023}
}

@article{Ducrot2025,
archivePrefix = {arXiv},
arxivId = {2412.11627},
author = {Ducrot, Elsa and Lagage, Pierre Olivier and Min, Michiel and Gillon, Micha{\"{e}}l and Bell, Taylor J. and Tremblin, Pascal and Greene, Thomas and Dyrek, Achr{\`{e}}ne and Bouwman, Jeroen and Waters, Rens and G{\"{u}}del, Manuel and Henning, Thomas and Vandenbussche, Bart and Absil, Olivier and Barrado, David and Boccaletti, Anthony and Coulais, Alain and Decin, Leen and Edwards, Billy and Gastaud, Ren{\'{e}} and Glasse, Alistair and Kendrew, Sarah and Olofsson, Goran and Patapis, Polychronis and Pye, John and Rouan, Daniel and Whiteford, Niall and Argyriou, Ioannis and Cossou, Christophe and Glauser, Adrian M. and Krause, Oliver and Lahuis, Fred and Royer, Pierre and Scheithauer, Silvia and Colina, Luis and van Dishoeck, Ewine F. and Ostlin, G{\"{o}}ran and Ray, Tom P. and Wright, Gillian},
doi = {10.1038/s41550-024-02428-z},
eprint = {2412.11627},
issn = {23973366},
journal = {Nature Astronomy},
number = {3},
pages = {358--369},
title = {{Combined analysis of the 12.8 and 15 $\mu$m JWST/MIRI eclipse observations of TRAPPIST-1 b}},
volume = {9},
year = {2025}
}

@article{Kuchner2003,
archivePrefix = {arXiv},
arxivId = {astro-ph/0303186},
author = {Kuchner, Marc J.},
doi = {10.1086/378397},
eprint = {0303186},
issn = {0004-637X},
journal = {The Astrophysical Journal},
number = {1},
pages = {L105--L108},
primaryClass = {astro-ph},
title = {{Volatile-rich Earth-Mass Planets in the Habitable Zone}},
volume = {596},
year = {2003}
}

@article{Leger2004,
title = {A new family of planets? “Ocean-Planets”},
journal = {Icarus},
volume = {169},
number = {2},
pages = {499-504},
year = {2004},
issn = {0019-1035},
doi = {https://doi.org/10.1016/j.icarus.2004.01.001},
url = {https://www.sciencedirect.com/science/article/pii/S001910350400020X},
author = {A. Léger and F. Selsis and C. Sotin and T. Guillot and D. Despois and D. Mawet and M. Ollivier and A. Labèque and C. Valette and F. Brachet and B. Chazelas and H. Lammer}
}

@incollection{Lichtenberg2025,
title = {Super-Earths and Earth-like exoplanets},
editor = {Ariel Anbar and Dominique Weis},
booktitle = {Treatise on Geochemistry (Third edition)},
publisher = {Elsevier},
edition = {Third edition},
address = {Oxford},
pages = {51-112},
year = {2025},
isbn = {978-0-323-99763-8},
doi = {https://doi.org/10.1016/B978-0-323-99762-1.00122-4},
url = {https://www.sciencedirect.com/science/article/pii/B9780323997621001224},
author = {Tim Lichtenberg and Yamila Miguel}
}

@misc{Nixon2025,
	title={Magma ocean interactions can explain JWST observations of the sub-Neptune TOI-270 d}, 
	author={Matthew C. Nixon and R. Sander Somers and Arjun B. Savel and Jegug Ih and Eliza M. -R. Kempton and Edward D. Young and Hilke E. Schlichting and Tim Lichtenberg and Luis Welbanks and William Misener and Anjali A. A. Piette and Nicholas F. Wogan},
	year={2025},
	eprint={2510.07367},
	archivePrefix={arXiv},
	primaryClass={astro-ph.EP},
	url={https://arxiv.org/abs/2510.07367}, 
}

@article{Seo2024,
archivePrefix = {arXiv},
arxivId = {2408.17056},
author = {Seo, Chanoul and Ito, Yuichi and Fujii, Yuka},
doi = {10.3847/1538-4357/ad7461},
eprint = {2408.17056},
issn = {0004-637X},
journal = {The Astrophysical Journal},
number = {1},
pages = {14},
publisher = {IOP Publishing},
title = {{Role of Magma Oceans in Controlling Carbon and Oxygen of Sub-Neptune Atmospheres}},
volume = {975},
year = {2024}
}

@article{Werlen2025,
archivePrefix = {arXiv},
arxivId = {2504.20450},
author = {Werlen, Aaron and Dorn, Caroline and Schlichting, Hilke E. and Grimm, Simon L. and Young, Edward D.},
doi = {10.3847/2041-8213/adf185},
eprint = {2504.20450},
issn = {2041-8205},
journal = {The Astrophysical Journal Letters},
number = {2},
pages = {L55},
title = {{Atmospheric C/O Ratios of Sub-Neptunes with Magma Oceans: Homemade rather than Inherited}},
volume = {988},
year = {2025}
}

@article{Dorn2021,
	archivePrefix = {arXiv},
	arxivId = {2110.15069},
	author = {Dorn, Caroline and Lichtenberg, Tim},
	doi = {10.3847/2041-8213/ac33af},
	eprint = {2110.15069},
	issn = {2041-8205},
	journal = {The Astrophysical Journal Letters},
	month = {nov},
	number = {1},
	pages = {L4},
	publisher = {IOP Publishing},
	title = {{Hidden Water in Magma Ocean Exoplanets}},
	url = {http://dx.doi.org/10.3847/2041-8213/ac33af https://iopscience.iop.org/article/10.3847/2041-8213/ac33af},
	volume = {922},
	year = {2021}
}

@article{melosh2007hydrocode,
  title={A hydrocode equation of state for SiO2},
  author={Melosh, HJ},
  journal={Meteoritics \& planetary science},
  volume={42},
  number={12},
  pages={2079--2098},
  year={2007},
  publisher={Wiley Online Library}
}

@article{connolly2009geodynamic,
  title={The geodynamic equation of state: what and how},
  author={Connolly, JAD},
  journal={Geochemistry, geophysics, geosystems},
  volume={10},
  number={10},
  year={2009},
  publisher={Wiley Online Library}
}

@article{gialluca2024implications,
  title={The Implications of Thermal Hydrodynamic Atmospheric Escape on the TRAPPIST-1 Planets},
  author={Gialluca, Megan T and Barnes, Rory and Meadows, Victoria S and Garcia, Rodolfo and Birky, Jessica and Agol, Eric},
  journal={The Planetary Science Journal},
  volume={5},
  number={6},
  pages={137},
  year={2024},
  publisher={IOP Publishing}
}

@article{krissansen2023implications,
  title={Implications of atmospheric nondetections for TRAPPIST-1 inner planets on atmospheric retention prospects for outer planets},
  author={Krissansen-Totton, Joshua},
  journal={The Astrophysical Journal Letters},
  volume={951},
  number={2},
  pages={L39},
  year={2023},
  publisher={IOP Publishing}
}

@article{pass2025receding,
  title={The Receding Cosmic Shoreline of Mid-to-late M Dwarfs: Measurements of Active Lifetimes Worsen Challenges for Atmosphere Retention by Rocky Exoplanets},
  author={Pass, Emily K and Charbonneau, David and Vanderburg, Andrew},
  journal={The Astrophysical Journal Letters},
  volume={986},
  number={1},
  pages={L3},
  year={2025},
  publisher={IOP Publishing}
}

@article{van2025habitable,
  title={Habitable Zone and Atmosphere Retention Distance (HaZARD)-Stellar-evolution-dependent loss models of secondary atmospheres},
  author={Van Looveren, Gwena{\"e}l and Saikia, Sudeshna Boro and Herbort, Oliver and Schleich, Simon and G{\"u}del, Manuel and Johnstone, Colin and Kislyakova, Kristina},
  journal={Astronomy \& Astrophysics},
  volume={694},
  pages={A310},
  year={2025},
  publisher={EDP Sciences}
}

@article{acuna2021characterisation,
  title={Characterisation of the hydrospheres of TRAPPIST-1 planets},
  author={Acu{\~n}a, Lorena and Deleuil, Magali and Mousis, Olivier and Marcq, Emmanuel and Levesque, Ma{\"e}va and Aguichine, Artyom},
  journal={Astronomy \& Astrophysics},
  volume={647},
  pages={A53},
  year={2021},
  publisher={EDP Sciences}
}

@article{JiXuan_2025,
	doi = {10.3847/1538-4357/adfe69},
	url = {https://doi.org/10.3847/1538-4357/adfe69},
	year = {2025},
	month = {oct},
	publisher = {The American Astronomical Society},
	volume = {992},
	number = {2},
	pages = {198},
	author = {Ji, Xuan and Chatterjee, Richard D. and Coy, Brandon Park and Kite, Edwin},
	title = {The Cosmic Shoreline Revisited: A Metric for Atmospheric Retention Informed by Hydrodynamic Escape},
	journal = {The Astrophysical Journal}
}

@article{Bower_2025,
	title={Diversity of Low-mass Planet Atmospheres in the C–H–O–N–S–Cl System with Interior Dissolution, Nonideality, and Condensation: Application to TRAPPIST-1e and Sub-Neptunes},
	volume={995},
	ISSN={1538-4357},
	url={http://dx.doi.org/10.3847/1538-4357/ae1479},
	DOI={10.3847/1538-4357/ae1479},
	number={1},
	journal={The Astrophysical Journal},
	publisher={American Astronomical Society},
	author={Bower, Dan J. and Thompson, Maggie A. and Hakim, Kaustubh and Tian, Meng and Sossi, Paolo A.},
	year={2025},
	month=dec, pages={59} }

@misc{Chatterjee2025,
	title={Novel Physics of Escaping Secondary Atmospheres May Shape the Cosmic Shoreline}, 
	author={Richard D. Chatterjee and Raymond T. Pierrehumbert},
	year={2025},
	eprint={2412.05188},
	archivePrefix={arXiv},
	primaryClass={astro-ph.EP},
	url={https://arxiv.org/abs/2412.05188}, 
}

@article{Fleming_2020,
	doi = {10.3847/1538-4357/ab77ad},
	url = {https://doi.org/10.3847/1538-4357/ab77ad},
	year = {2020},
	month = {mar},
	publisher = {The American Astronomical Society},
	volume = {891},
	number = {2},
	pages = {155},
	author = {Fleming, David P. and Barnes, Rory and Luger, Rodrigo and VanderPlas, Jacob T.},
	title = {On the XUV Luminosity Evolution of TRAPPIST-1},
	journal = {The Astrophysical Journal}
}

@article{Armstrong2019,
	author = {Armstrong, Katherine and Frost, Daniel J. and McCammon, Catherine A. and Rubie, David C. and Ballaran, Tiziana Boffa},
	doi = {10.1126/science.aax8376},
	issn = {10959203},
	journal = {Science},
	number = {6456},
	pages = {903--906},
	pmid = {31467218},
	title = {{Deep magma ocean formation set the oxidation state of Earth's mantle}},
	volume = {365},
	year = {2019}
}

@article{Luger2017,
	archivePrefix = {arXiv},
	arxivId = {1703.04166},
	author = {Luger, Rodrigo and Sestovic, Marko and Kruse, Ethan and Grimm, Simon L. and Demory, Brice Olivier and Agol, Eric and Bolmont, Emeline and Fabrycky, Daniel and Fernandes, Catarina S. and {Van Grootel}, Val{\'{e}}rie and Burgasser, Adam and Gillon, Micha{\"{e}}l and Ingalls, James G. and Jehin, Emmanu{\"{e}}l and Raymond, Sean N. and Selsis, Franck and Triaud, Amaury H.M.J. and Barclay, Thomas and Barentsen, Geert and Howell, Steve B. and Delrez, Laetitia and {De Wit}, Julien and Foreman-Mackey, Daniel and Holdsworth, Daniel L. and Leconte, J{\'{e}}r{\'{e}}my and Lederer, Susan and Turbet, Martin and Almleaky, Yaseen and Benkhaldoun, Zouhair and Magain, Pierre and Morris, Brett M. and Heng, Kevin and Queloz, DIdier},
	doi = {10.1038/s41550-017-0129},
	eprint = {1703.04166},
	issn = {23973366},
	journal = {Nature Astronomy},
	number = {May},
	pages = {1--8},
	title = {{A seven-planet resonant chain in TRAPPIST-1}},
	volume = {1},
	year = {2017}
}

@article{Piaulet-Ghorayeb2025,
	archivePrefix = {arXiv},
	arxivId = {2508.08416},
	author = {Piaulet-Ghorayeb, Caroline and Benneke, Bj{\"{o}}rn and Turbet, Martin and Moore, Keavin and Roy, Pierre-Alexis and Lim, Olivia and Doyon, Ren{\'{e}} and Fauchez, Thomas J. and Albert, Lo{\"{i}}c and Radica, Michael and Coulombe, Louis-Philippe and Lafreni{\`{e}}re, David and Cowan, Nicolas B. and Belzile, Danika and Musfirat, Kamrul and Kaur, Mehramat and L'Heureux, Alexandrine and Johnstone, Doug and MacDonald, Ryan J. and Allart, Romain and Dang, Lisa and Kaltenegger, Lisa and Pelletier, Stefan and Rowe, Jason F. and Taylor, Jake and Turner, Jake D.},
	doi = {10.3847/1538-4357/adf207},
	eprint = {2508.08416},
	issn = {1538-4357},
	journal = {The Astrophysical Journal},
	number = {2},
	pages = {0},
	publisher = {IOP Publishing},
	title = {{Strict limits on potential secondary atmospheres on the temperate rocky exo-Earth TRAPPIST-1 d}},
	url = {http://arxiv.org/abs/2508.08416},
	volume = {989},
	year = {2025}
}

@article{Agol2021,
	archivePrefix = {arXiv},
	arxivId = {2010.01074},
	author = {Agol, Eric and Dorn, Caroline and Grimm, Simon L. and Turbet, Martin and Ducrot, Elsa and Delrez, Laetitia and Gillon, Micha{\"{e}}l and Demory, Brice Olivier and Burdanov, Artem and Barkaoui, Khalid and Benkhaldoun, Zouhair and Bolmont, Emeline and Burgasser, Adam and Carey, Sean and {De Wit}, Julien and Fabrycky, Daniel and Foreman-Mackey, Daniel and Haldemann, Jonas and Hernandez, David M. and Ingalls, James and Jehin, Emmanuel and Langford, Zachary and Leconte, J{\'{e}}r{\'{e}}my and Lederer, Susan M. and Luger, Rodrigo and Malhotra, Renu and Meadows, Victoria S. and Morris, Brett M. and Pozuelos, Francisco J. and Queloz, Didier and Raymond, Sean N. and Selsis, Franck and Sestovic, Marko and Triaud, Amaury H.M.J. and {Van Grootel}, Valerie},
	doi = {10.3847/PSJ/abd022},
	eprint = {2010.01074},
	issn = {26323338},
	journal = {Planetary Science Journal},
	number = {1},
	pages = {1},
	publisher = {IOP Publishing},
	title = {{Refining the transit-timing and photometric analysis of TRAPPIST-1: Masses, Radii, densities, dynamics, and ephemerides}},
	url = {http://dx.doi.org/10.3847/PSJ/abd022},
	volume = {2},
	year = {2021}
}

@article{Unterborn2018,
	archivePrefix = {arXiv},
	arxivId = {1706.02689},
	author = {Unterborn, Cayman T. and Desch, Steven J. and Hinkel, Natalie R. and Lorenzo, Alejandro},
	doi = {10.1038/s41550-018-0411-6},
	eprint = {1706.02689},
	issn = {23973366},
	journal = {Nature Astronomy},
	number = {4},
	pages = {297--302},
	title = {{Inward migration of the TRAPPIST-1 planets as inferred from their water-rich compositions}},
	volume = {2},
	year = {2018}
}

@article{Grimm2018,
	author = {Grimm, Simon L and Demory, Brice-olivier and Gillon, Micha{\"{e}}l and Dorn, Caroline and Agol, Eric and Burdanov, Artem and Delrez, Laetitia and Sestovic, Marko and Triaud, Amaury H M J and Turbet, Martin and Bolmont, {\'{E}}meline and Caldas, Anthony and de Wit, Julien and Jehin, Emmanu{\"{e}}l and Leconte, J{\'{e}}r{\'{e}}my and Raymond, Sean N. and {Van Grootel}, Val{\'{e}}rie and Burgasser, Adam J. and Carey, Sean and Fabrycky, Daniel and Heng, Kevin and Hernandez, David M and Ingalls, James G and Lederer, Susan and Selsis, Franck and Queloz, Didier},
	doi = {10.1051/0004-6361/201732233},
	issn = {0004-6361},
	journal = {Astronomy \& Astrophysics},
	month = {may},
	pages = {A68},
	title = {{The nature of the TRAPPIST-1 exoplanets}},
	url = {https://www.aanda.org/10.1051/0004-6361/201732233},
	volume = {613},
	year = {2018}
}

@article{Gillon2017,
	author = {Gillon, Micha{\"{e}}l and Triaud, Amaury H.M.J. and Demory, Brice Olivier and Jehin, Emmanu{\"{e}}l and Agol, Eric and Deck, Katherine M. and Lederer, Susan M. and {De Wit}, Julien and Burdanov, Artem and Ingalls, James G. and Bolmont, Emeline and Leconte, Jeremy and Raymond, Sean N. and Selsis, Franck and Turbet, Martin and Barkaoui, Khalid and Burgasser, Adam and Burleigh, Matthew R. and Carey, Sean J. and Chaushev, Aleksander and Copperwheat, Chris M. and Delrez, Laetitia and Fernandes, Catarina S. and Holdsworth, Daniel L. and Kotze, Enrico J. and {Van Grootel}, Val{\'{e}}rie and Almleaky, Yaseen and Benkhaldoun, Zouhair and Magain, Pierre and Queloz, Didier},
	doi = {10.1038/nature21360},
	issn = {14764687},
	journal = {Nature},
	number = {7642},
	pages = {456--460},
	pmid = {28230125},
	publisher = {Nature Publishing Group},
	title = {{Seven temperate terrestrial planets around the nearby ultracool dwarf star TRAPPIST-1}},
	url = {http://dx.doi.org/10.1038/nature21360},
	volume = {542},
	year = {2017}
}

@article{Quarles2017,
	archivePrefix = {arXiv},
	arxivId = {1704.02261},
	author = {Quarles, B. and Quintana, E. V. and Lopez, E. and Schlieder, J. E. and Barclay, T.},
	doi = {10.3847/2041-8213/aa74bf},
	eprint = {1704.02261},
	issn = {2041-8205},
	journal = {The Astrophysical Journal Letters},
	number = {1},
	pages = {L5},
	publisher = {IOP Publishing},
	title = {{Plausible Compositions of the Seven TRAPPIST-1 Planets Using Long-term Dynamical Simulations}},
	url = {http://dx.doi.org/10.3847/2041-8213/aa74bf},
	volume = {842},
	year = {2017}
}

@article{ElkinsTanton2008,
	author = {Elkins‐Tanton, Linda T. and Seager, Sara},
	doi = {10.1086/592316},
	issn = {0004-637X},
	journal = {The Astrophysical Journal},
	number = {1},
	pages = {628--635},
	title = {{Coreless Terrestrial Exoplanets}},
	volume = {688},
	year = {2008}
}

@article{Krissansen-Totton2022,
	archivePrefix = {arXiv},
	arxivId = {2207.04164},
	author = {Krissansen-Totton, J. and Fortney, J. J.},
	doi = {10.3847/1538-4357/ac69cb},
	eprint = {2207.04164},
	issn = {0004-637X},
	journal = {The Astrophysical Journal},
	number = {1},
	pages = {115},
	publisher = {IOP Publishing},
	title = {{Predictions for Observable Atmospheres of Trappist-1 Planets from a Fully Coupled Atmosphere–Interior Evolution Model}},
	url = {http://dx.doi.org/10.3847/1538-4357/ac69cb},
	volume = {933},
	year = {2022}
}

@article{Seager2007,
	archivePrefix = {arXiv},
	arxivId = {0707.2895},
	author = {Seager, S. and Kuchner, M. and Hier‐Majumder, C. A. and Militzer, B.},
	doi = {10.1086/521346},
	eprint = {0707.2895},
	issn = {0004-637X},
	journal = {The Astrophysical Journal},
	number = {2},
	pages = {1279--1297},
	title = {{Mass‐Radius Relationships for Solid Exoplanets}},
	volume = {669},
	year = {2007}
}

@article{Spaargaren2023,
	archivePrefix = {arXiv},
	arxivId = {2211.01800},
	author = {Spaargaren, Rob J. and Wang, Haiyang S. and Mojzsis, Stephen J and Ballmer, Maxim D and Tackley, Paul J},
	doi = {10.3847/1538-4357/acac7d},
	eprint = {2211.01800},
	issn = {0004-637X},
	journal = {The Astrophysical Journal},
	month = {may},
	number = {1},
	pages = {53},
	title = {{Plausible Constraints on the Range of Bulk Terrestrial Exoplanet Compositions in the Solar Neighborhood}},
	url = {http://arxiv.org/abs/2211.01800 http://dx.doi.org/10.3847/1538-4357/acac7d https://iopscience.iop.org/article/10.3847/1538-4357/acac7d},
	volume = {948},
	year = {2023}
}

@article{Dorn2015,
	author = {Dorn, Caroline and Khan, Amir and Heng, Kevin and Connolly, James A D and Alibert, Yann and Benz, Willy and Tackley, Paul},
	doi = {10.1051/0004-6361/201424915},
	issn = {0004-6361},
	journal = {Astronomy \& Astrophysics},
	month = {may},
	pages = {A83},
	title = {{Can we constrain the interior structure of rocky exoplanets from mass and radius measurements?}},
	url = {http://www.aanda.org/10.1051/0004-6361/201424915},
	volume = {577},
	year = {2015}
}

@article{Noack2020,
	author = {Noack, Lena and Lasbleis, Marine},
	doi = {10.1051/0004-6361/202037723},
	issn = {0004-6361},
	journal = {Astronomy \& Astrophysics},
	month = {jun},
	pages = {A129},
	title = {{Parameterisations of interior properties of rocky planets}},
	url = {https://www.aanda.org/10.1051/0004-6361/202037723},
	volume = {638},
	year = {2020}
}

@article{Zeng2019,
	author = {Zeng, Li and Jacobsen, Stein B. and Sasselov, Dimitar D. and Petaev, Michail I. and Vanderburg, Andrew and Lopez-Morales, Mercedes and Perez-Mercader, Juan and Mattsson, Thomas R. and Li, Gongjie and Heising, Matthew Z. and Bonomo, Aldo S. and Damasso, Mario and Berger, Travis A. and Cao, Hao and Levi, Amit and Wordsworth, Robin D.},
	doi = {10.1073/pnas.1812905116},
	issn = {10916490},
	journal = {Proceedings of the National Academy of Sciences of the United States of America},
	number = {20},
	pages = {9723--9728},
	pmid = {31036661},
	title = {{Growth model interpretation of planet size distribution}},
	volume = {116},
	year = {2019}
}

@article{Wang2018,
	archivePrefix = {arXiv},
	arxivId = {1708.08718},
	author = {Wang, Haiyang S. and Lineweaver, Charles H. and Ireland, Trevor R.},
	doi = {10.1016/j.icarus.2017.08.024},
	eprint = {1708.08718},
	issn = {00191035},
	journal = {Icarus},
	month = {jan},
	pages = {460--474},
	publisher = {Elsevier Inc.},
	title = {{The elemental abundances (with uncertainties) of the most Earth-like planet}},
	url = {http://dx.doi.org/10.1016/j.icarus.2017.08.024 https://linkinghub.elsevier.com/retrieve/pii/S0019103517302221},
	volume = {299},
	year = {2018}
}

@article{Luo2024,
	author = {Luo, Haiyang and Dorn, Caroline and Deng, Jie},
	doi = {10.1038/s41550-024-02347-z},
	issn = {23973366},
	journal = {Nature Astronomy},
	number = {November},
	publisher = {Springer US},
	title = {{The interior as the dominant water reservoir in super-Earths and sub-Neptunes}},
	url = {http://dx.doi.org/10.1038/s41550-024-02347-z},
	volume = {8},
	year = {2024}
}

@article{Luque2022,
	archivePrefix = {arXiv},
	arxivId = {2209.03871},
	author = {Luque, Rafael and Pall{\'{e}}, Enric},
	doi = {10.1126/science.abl7164},
	eprint = {2209.03871},
	issn = {10959203},
	journal = {Science},
	number = {6611},
	pages = {1211--1214},
	pmid = {36074855},
	title = {{Density, not radius, separates rocky and water-rich small planets orbiting M dwarf stars}},
	volume = {377},
	year = {2022}
}

@article{Valencia2007,
	archivePrefix = {arXiv},
	arxivId = {0704.3454},
	author = {Valencia, Diana and Sasselov, Dimitar D. and O'Connell, Richard J.},
	doi = {10.1086/519554},
	eprint = {0704.3454},
	issn = {0004-637X},
	journal = {The Astrophysical Journal},
	number = {2},
	pages = {1413--1420},
	title = {{Detailed Models of Super‐Earths: How Well Can We Infer Bulk Properties?}},
	volume = {665},
	year = {2007}
}

@article{Dorn2018,
	archivePrefix = {arXiv},
	arxivId = {1808.01803},
	author = {Dorn, Caroline and Mosegaard, Klaus and Grimm, Simon L. and Alibert, Yann},
	doi = {10.3847/1538-4357/aad95d},
	eprint = {1808.01803},
	issn = {0004-637X},
	journal = {The Astrophysical Journal},
	number = {1},
	pages = {20},
	publisher = {IOP Publishing},
	title = {{Interior Characterization in Multiplanetary Systems: TRAPPIST-1}},
	url = {http://dx.doi.org/10.3847/1538-4357/aad95d},
	volume = {865},
	year = {2018}
}

@article{Bouhifd2011,
	author = {Bouhifd, M. A. and Jephcoat, A. P.},
	doi = {10.1016/j.epsl.2011.05.006},
	issn = {0012821X},
	journal = {Earth and Planetary Science Letters},
	month = {jul},
	number = {3-4},
	pages = {341--348},
	title = {{Convergence of Ni and Co metal-silicate partition coefficients in the deep magma-ocean and coupled silicon-oxygen solubility in iron melts at high pressures}},
	volume = {307},
	year = {2011}
}

@article{Siebert2012,
	author = {Siebert, Julien and Badro, James and Antonangeli, Daniele and Ryerson, Frederick J.},
	doi = {10.1016/j.epsl.2012.01.013},
	issn = {0012821X},
	journal = {Earth and Planetary Science Letters},
	month = {mar},
	pages = {189--197},
	publisher = {Elsevier B.V.},
	title = {{Metal-silicate partitioning of Ni and Co in a deep magma ocean}},
	volume = {321-322},
	year = {2012}
}

@article{Fischer2020,
	author = {Fischer, Rebecca A. and Cottrell, Elizabeth and Hauri, Erik and Lee, Kanani K.M. and {Le Voyer}, Marion and {M Lee}, Kanani K and {Le Voyer}, Marion},
	doi = {10.1073/pnas.1919930117},
	issn = {10916490},
	journal = {Proceedings of the National Academy of Sciences of the United States of America},
	number = {16},
	pages = {8743--8749},
	pmid = {32229562},
	title = {{The carbon content of Earth and its core}},
	url = {www.pnas.org/cgi/doi/10.1073/pnas.1919930117},
	volume = {117},
	year = {2020}
}

@article{Huang2024,
	author = {Huang, Dongyang and Siebert, Julien and Sossi, Paolo and Kubik, Edith and Avice, Guillaume and Murakami, Motohiko},
	doi = {10.1016/j.gca.2024.05.010},
	issn = {00167037},
	journal = {Geochimica et Cosmochimica Acta},
	month = {jul},
	number = {May},
	pages = {100--112},
	publisher = {Elsevier Ltd},
	title = {{Nitrogen sequestration in the core at megabar pressure and implications for terrestrial accretion}},
	url = {https://linkinghub.elsevier.com/retrieve/pii/S0016703724002321},
	volume = {376},
	year = {2024}
}

@article{Lodders2003,
	archivePrefix = {arXiv},
	arxivId = {1409.7398},
	author = {Lodders, Katharina},
	doi = {10.1086/375492},
	eprint = {1409.7398},
	isbn = {0004-637X},
	issn = {0004-637X},
	journal = {The Astrophysical Journal},
	month = {jul},
	number = {2},
	pages = {1220--1247},
	pmid = {24553238},
	title = {{Solar System Abundances and Condensation Temperatures of the Elements}},
	url = {https://iopscience.iop.org/article/10.1086/375492},
	volume = {591},
	year = {2003}
}

@article{Siebert2013,
	author = {Siebert, Julien and Badro, James and Antonangeli, Daniele and Ryerson, Frederick J},
	doi = {10.1126/science.1227923},
	isbn = {2011014204},
	journal = {Science},
	number = {6124},
	pages = {1194--1197},
	title = {{Terrestrial Accretion Under Oxidizing Conditions}},
	volume = {339},
	year = {2013}
}

@article{Huang2020,
	author = {Huang, Dongyang and Badro, James and Siebert, Julien},
	doi = {10.1073/pnas.2007982117},
	issn = {0027-8424},
	journal = {Proceedings of the National Academy of Sciences},
	month = {nov},
	number = {45},
	pages = {27893--27898},
	publisher = {National Academy of Sciences},
	title = {{The niobium and tantalum concentration in the mantle constrains the composition of Earth's primordial magma ocean}},
	url = {http://www.pnas.org/lookup/doi/10.1073/pnas.2007982117 https://www.pnas.org/content/early/2020/10/22/2007982117 https://pnas.org/doi/full/10.1073/pnas.2007982117},
	volume = {117},
	year = {2020}
}

@article{Huang2021,
	author = {Huang, Dongyang and Siebert, Julien and Badro, James},
	doi = {10.1016/j.gca.2021.06.031},
	issn = {00167037},
	journal = {Geochimica et Cosmochimica Acta},
	month = {oct},
	pages = {19--31},
	publisher = {Elsevier Ltd},
	title = {{High pressure partitioning behavior of Mo and W and late sulfur delivery during Earth's core formation}},
	url = {https://www.sciencedirect.com/science/article/pii/S0016703721003896 https://linkinghub.elsevier.com/retrieve/pii/S0016703721003896},
	volume = {310},
	year = {2021}
}

@misc{young2025,
      title={Differentiation, the exception not the rule -- Evidence for full miscibility in sub-Neptune interiors}, 
      author={Edward D. Young and Aaron Werlen and Sarah P. Marcum and Lars Stixrude and Cornelis P. Dullemond},
      year={2025},
      eprint={2507.00947},
      archivePrefix={arXiv},
      primaryClass={astro-ph.EP},
      url={https://arxiv.org/abs/2507.00947}, 
}
\bibliographystyle{aasjournalv7}



\end{document}